\documentclass{IEEEtran}
\usepackage{amsmath,amsfonts}
\usepackage{algorithmic}
\usepackage{array}
\usepackage[caption=false,font=normalsize,labelfont=sf,textfont=sf]{subfig}
\usepackage{textcomp}
\usepackage{stfloats}
\usepackage{url}
\usepackage{verbatim}
\usepackage{graphicx}
\usepackage{xcolor}
\usepackage{units}

\usepackage{biblatex}
\title{Effects of Social Interactions in Self-Organising Railway Traffic Management}
\author{Fabio Oddi, Federico Naldini, Leo D'Amato, Gr{\'e}gory Marli{\`e}re, Paola Pellegrini, Vito Trianni
\thanks{Fabio Oddi, Leo D'Amato, Vito Trianni are with the Institute for Cognitive Sciences and Technologies, CNR, Via Giandomenico Romagnosi 18A, 00196, Rome, Italy.} 
\thanks{Federico Naldini, Gr{\'e}gory Marli{\`e}re and Paola Pellegrini are with the Univ. Gustave Eiffel, COSYS-ESTAS, F-59650 Villeneuve d'Ascq, France.}
\thanks{Fabio Oddi is with Dipartimento di Ingegneria Informatica, Automatica e Gestionale, La Sapienza Universit{\'a} di Roma, Via Ariosto 25, 00185, Rome, Italy.}
\thanks{Leo D'Amato is with the Dipartimento di Automatica e Informatica, Politecnico di Torino, Corso Duca degli Abruzzi, 24, 10129 Torino, Italy.}
}
\date{}
\begin{document}
\maketitle

\begin{abstract}
Recent research is exploring self-organised traffic management as a solution for scaling to complex real-world networks. In such a system, trains predict their neighbourhood, produce traffic plan hypotheses, and agree via consensus with neighbours on a future traffic plan to be implemented. This paper investigates a structural parameter within this pipeline: the predictive neighbourhood horizon. The horizon is used by trains to identify future potential conflicts with neighbours, and to establish the local interaction topology, that is, the subset of trains to negotiate with. As the primary design variable, the horizon directly determines the size and density of the social interaction graph, whereas its impact on the complexity of local sub-problems and the distributed consensus dynamics represents a trade-off to be explored. Through a closed-loop simulation framework the study evaluates how variations of the horizon impact the overall decentralised coordination process, from initial conflict detection to distributed schedule consensus. The analysis focuses on investigating the potential trade-off introduced by the horizon choice: balancing local tractability and computational responsiveness with the need for global schedule coherence and feasibility in safety-critical environments. Contrary to intuition, our empirical results indicate that the short time horizons suffice, while long values compromise local tractability and computational responsiveness with no gain in global schedule optimality.
\end{abstract}

\section{Introduction}\label{sec:intro}
Real‑time railway traffic management has long relied on centralised dispatching, where a small number of decision centres resolve conflicts and limit delay propagation. To support decision making, research has proposed several optimisation tools~\cite{corman2015review,cacchiani2014overview,fang2015survey} or, more recently, machine learning techniques~\cite{ghasempour2020adaptive,khadilkar2018scalable,zhang2020fast}. While centralised approaches can be highly effective, they encounter well‑known limitations: in dense networks, computation times grow, scalability becomes a bottleneck, and practical deployment clashes with the need to integrate heterogeneous or commercially-sensitive information from competing railway undertakings. These constraints motivate a shift toward architectures that reduce dependence on a single decision authority and instead exploit local information and peer coordination to achieve timely, robust responses to unplanned events or the need to modify the route and the schedule.
Self‑organising traffic management systems (SO-TMS) reframe the control problem by delegating decision authority to local agents that make context‑aware choices and coordinate only with nearby traffic~\cite{shang2018distributed,vanthielen2019conflict,yong2017swarm}. This paradigm reduces the scope of each decision problem, lowers communication overhead and preserves operator privacy by keeping sensitive information local.

Self‑organisation is a recurring paradigm in natural and engineered systems, where global coordination emerges from local interactions among autonomous agents. Biological examples such as ant colonies and honeybee swarms achieve complex collective behaviours without centralised control~\cite{detrain2006self,seeley2012stop}, and artificial multi‑agent systems—robot swarms, cognitive radio networks, and Internet‑of‑Things infrastructures—exploit simple local rules to produce coherent global outcomes~\cite{dorigo2021swarm,liu2023cognitive,arellanes2021self}. In transportation, analogous phenomena are observed at multiple scales: ant trails self‑regulate to reduce congestion~\cite{strombom2018self}, pedestrian crowds form lanes and adapt flow patterns~\cite{moussaid2012traffic,murakami2021mutual}, and self‑organising traffic lights respond to local demand~\cite{helbing2005self,gershenson2012self}. These examples motivate the application of self-organised control principles to mobility systems, where locality can improve scalability and robustness relative to monolithic, centralised planners~\cite{bucchiarone2020agent,gerrits2024self,mittal2024efficient}.

In previous work, we introduced a modular architecture for SO-TMS that decomposes the decision process into a sequence of interoperable local operations~\cite{damato2020towards}: prediction, negotiation, conflict resolution and execution. A closed‑loop implementation of the SO‑TMS has been experimentally validated in recent work~\cite{naldini2026self}, showing that self‑organising strategies can match or even outperform state‑of‑the‑art centralised optimisation such as RECIFE‑MILP~\cite{pellegrini2015recife}. By narrowing the decision scope and enabling local recovery actions, the SO‑TMS can mitigate delay propagation and improve service reliability in scenarios where centralised solvers struggle to provide adequate solutions. At the same time, local decision‑making enables competing operators to participate without exposing commercially sensitive data, an important practical advantage for mixed‑traffic and open-access networks.

In this study, we focus our attention on a single structural choice in the SO-TMS pipeline: the temporal horizon ($T_h$) over which conflict prediction is conducted. This is a key parameter that determines the topology of the social interactions among neighbouring trains, which in turns has a bearing on all subsequent stages of the SO-TMS. 
Intuitively, the choice of $T_h$ regulates the balance between local planning and global optimality. Small values of $T_h$ could lead to easier but myopic planning, while large values should lead to more complex planning that however are closer to the global optimum. 
Understanding this trade-off is crucial for identifying the structural conditions that provide optimal working conditions. Our results demonstrate that, contrary to intuition, short time horizons are entirely sufficient. Extending the predictive window compromises computational responsiveness without providing any tangible gain in global schedule optimality.

The remainder of this paper is organised as follows: Section \ref{sec:adtm} introduces the SO-TMS architecture, while Section \ref{subsec:nhr} formalises the process leading to the definition of the train interaction network. Section \ref{sec:exp} details the experimental setup, Section \ref{sec:res} discusses the simulation results, and Section \ref{sec:concl} concludes the work with directions for future research.

\section{Self-Organising Traffic Management}\label{sec:adtm}
Railway traffic management poses particular challenges for decentralisation because of hard safety constraints, tight spatio-temporal coupling between services, and heterogeneous commercial objectives. Prior work showed that localised or distributed strategies can be effective in constrained contexts---such as metro lines, isolated junctions, or short-horizon conflict avoidance---but these studies stop short of proposing a general, operationally realistic architecture for mainline networks~\cite{shang2018distributed,yong2017swarm,vanthielen2019conflict}.
To bridge this gap, the proposed SO-TMS shifts the decision-making burden to the individual trains while preserving a central controller for the final generation of a global traffic management plan. This central step is limited to assembling decisions made by individual trains, and possible solving trailing conflicts in case of consensus failures or out-of-horizon long-term conflicts In this way, by decomposing the global scheduling problem and keeping individual decisions private, the framework allows competing operators to negotiate feasible paths dynamically without the need to disclose commercially sensitive data to a central controller. This decentralised approach is operationalised through a five-stages pipeline:
\begin{itemize}
    \item \textbf{Neighbourhood Identification:} As the foundational step, each train $t$ processes the current network state to detect potential conflicts over the time horizon $T_h$ and selects a specific set of interacting trains, defining its local neighbourhood $\mathcal{N}(t)$. This stage establishes the social interaction topology for all subsequent operations.
    
    \item \textbf{Hypothesis Generation and Sharing:} Using this defined topology, trains formulate candidate routing schedules (hypotheses) by solving a local Mixed Integer Linear Programming (MILP) problem. Each hypothesis incorporates the train's private objective function (e.g., estimated total delay). These generated hypotheses are then shared strictly within the train's specific neighbourhood $\mathcal{N}(t)$, and duplicate hypotheses are then discarded.
    
    \item \textbf{Hypothesis Compatibility:} Upon receiving proposals from neighbours, each train evaluates their pairwise compatibility. Two hypotheses are deemed incompatible if they allocate the same track section to different trains with overlapping time frames. This deterministic filtering produces a local mapping of mutually feasible plans.
    
    \item \textbf{Hypothesis Selection:} Based on the compatibility mapping, trains engage in a distributed negotiation mechanism inspired by voter models. Starting with their lowest-cost hypothesis, trains iteratively adapt their choices to align with neighbours, aiming to reach a local consensus on a schedule that maximises overall compatibility.
    
    \item \textbf{Merge:} In the final stage, selected hypotheses are collected centrally and merged to generate a global plan that guarantees global feasibility and strict compliance with safety constraints. At this step, targeted optimisations of the global schedule are applied to resolve any residual incompatibilities before the plan is executed.
\end{itemize}
As demonstrated in a closed-loop experimental analysis~\cite{naldini2026self}, the SO-TMS can mitigate delay propagation, yielding competitive results compared to state-of-the-art centralised optimisation methods such as RECIFE-MILP. These outcomes indicate that the framework successfully decomposes the global scheduling problem into tractable local sub-tasks, maintaining computational efficiency without compromising the overall coherence of the traffic plan. Furthermore, the results confirmed that the distributed consensus mechanism consistently produces feasible schedules, demonstrating the practical applicability of the approach in mixed-traffic environments where competing operators must coordinate without disclosing sensitive commercial data.

\section{Neighbourhood Identification}\label{subsec:nhr}
A central design choice in the SO-TMS pipeline is the predictive neighbourhood horizon $T_h$ used to identify trains that may get into conflict. The neighbourhood horizon governs the decomposition of the global problem into local sub-problems, hence a systematic analysis of the effects of this parameter is necessary to understand its cascading effects on the entire pipeline. This work isolates this parameter and evaluates its impact on computational effort, consensus stability, and operational performance in a realistic closed-loop setting.

To compute its neighbourhood, each train must estimate its own position and route alongside those of other trains. A route is defined by the sequence of \emph{Track Detection Sections (TDSs)}---i.e., portions of the physical railway where train presence can be detected---and the planned time interval of their utilisation. Neighbourhoods are updated at each iteration using two key data structures processed by the module:
\begin{itemize}
    \item \textbf{Real Time Traffic Plan (RTTP):} Represents the most recently updated schedule of trains over the infrastructure.
    \item \textbf{Traffic State Prediction (TSP):} A snapshot of the current network state at time $T$, provided by the simulator at a fixed rate (e.g., every 5 minutes). It includes the last known positions of the trains.
\end{itemize}
By combining these inputs, each train estimates the future track utilisations for each train, accounting for stopping, dwell, and departure times. Subsequently, each train $t$ defines its neighbourhood $\mathcal{N}(t)$ by identifying conflicting utilisation of TDSs within the time frame $[T, T+T_h]$. For each train $t$, let $[s_t(x), e_t(x)]$ be the predicted utilisation interval of the TDS $x$. Train $t'$ is included in the neighbourhood of train $t$ if there exists at least one TDS $x$ such that the predicted intervals overlap within the period $[T, T+T_h]$, i.e., $\mathcal{N}(t) = \{t' \mid \exists\ x : s_{t'}(x) < e_t(x) \lor s_t(x) < e_{t'}(x)\}$. Note that neighbourhood inclusion is reciprocal; that is, $t' \in \mathcal{N}(t) \iff t \in \mathcal{N}(t')$. Such mutual inclusion prevents divergent solutions.

Given the neighbourhood inclusion rule introduced above, we define the social interaction graph $G_{I}=(\mathcal{T},\mathcal{E})$, where $\mathcal{T}$ is the set of trains with a non-empty neighbourhood, and $\mathcal{E}$ denotes the set of edges. An edge connects two trains only if they are mutually neighbours. The choice of $T_h$ directly alters the topology of this graph, possibly inducing several operational trade-offs. A small $T_h$ yields sparse graphs with small connected components, likely shrinking the MILP search space and lowering communication demands. While that leads to easier optimisation, it could potentially increase the probability that relevant downstream interactions are omitted. Conversely, a large $T_h$ produces dense graphs and large components, increasing foresight and the likelihood of generating globally compatible hypotheses. However, this expands the combinatorial size of local hypothesis spaces, potentially reducing consensus convergence speed, which could negate the benefits of greater foresight under tight real-time constraints.

To validate these hypotheses with measurable outcomes, we define a set of indicators to evaluate how $T_h$ affects system performance. Topological indicators include the number and maximum size of connected components. Computational indicators encompass the MILP solve time, optimality distance, and hypothesis sharing increment. Finally, operational indicators include the total delay and the number of conflicts triggering a central repair. By sweeping $T_h$ and recording these metrics under controlled perturbations, we systematically quantify the operational shifts driven by the predictive horizon.

\section{Experimental Setup}\label{sec:exp}
We evaluate the effects of the neighbourhood horizon using a closed-loop simulation framework where the SO-TMS exchanges information with the OpenTrack~\cite{nash2004railroad} microscopic simulator at fixed five-minute intervals. At each step, the simulator provides a Traffic State Prediction (TSP) containing train positions and operational states, which is subsequently used to update the current RTTP. The initial  RTTP represents the plan before any perturbation takes place; thereafter, the RTTP produced as output from each pipeline iteration becomes the input for the subsequent round. Each train executes the SO-TMS pipeline and submits its selected hypothesis to a central integrator, which merges them into an updated RTTP for the simulator. This closed-loop arrangement captures the feedback dynamics between local decisions and overall traffic evolution.

Experiments are conducted on the 60\,km mixed-traffic corridor between Segrate and Ospitaletto (Italy), accommodating approximately 150 trains per day, including freight, intercity, regional, and high-speed services. To generate a high-density environment suitable for testing, the baseline timetable is compressed to minimise slack and path durations. Furthermore, scheduled train routes are partitioned into shorter, station-to-station segments linked by re-utilisation and platform compatibility constraints. Although this formulation increases the overall number of trains, it significantly bounds the combinatorial growth of alternative routes, accelerating the local MILP optimisation.

We evaluate the system performance across ten distinct perturbation scenarios. These are generated by sampling entrance delays from historical, type-dependent distributions. To emulate realistic disturbances, ten perturbed scenarios are generated by sampling entrance delays from historical, type-dependent distributions. The simulations commence at approximately 01:00 and conclude at 19:00, covering the full daily traffic cycle. Within this framework, we systematically vary $T_h$—including a 15-minute baseline—to measure its impact on the social interaction graph topology, hypothesis generation, consensus convergence, and the frequency of central merge repairs.

\section{Results}\label{sec:res}
The analysis of the neighbourhood horizon variations must be interpreted in light of the system's operational objectives: optimising infrastructure usage to reduce overall delays while guaranteeing compliance with safety constraints through centralised validation. To this end, the macroscopic impact on network performance is evaluated first, followed by a progressive analysis of the structural, computational, and negotiation behaviours that determine the global outcome.

\begin{figure}[ht]
    \centering
    \includegraphics[width=1\linewidth]{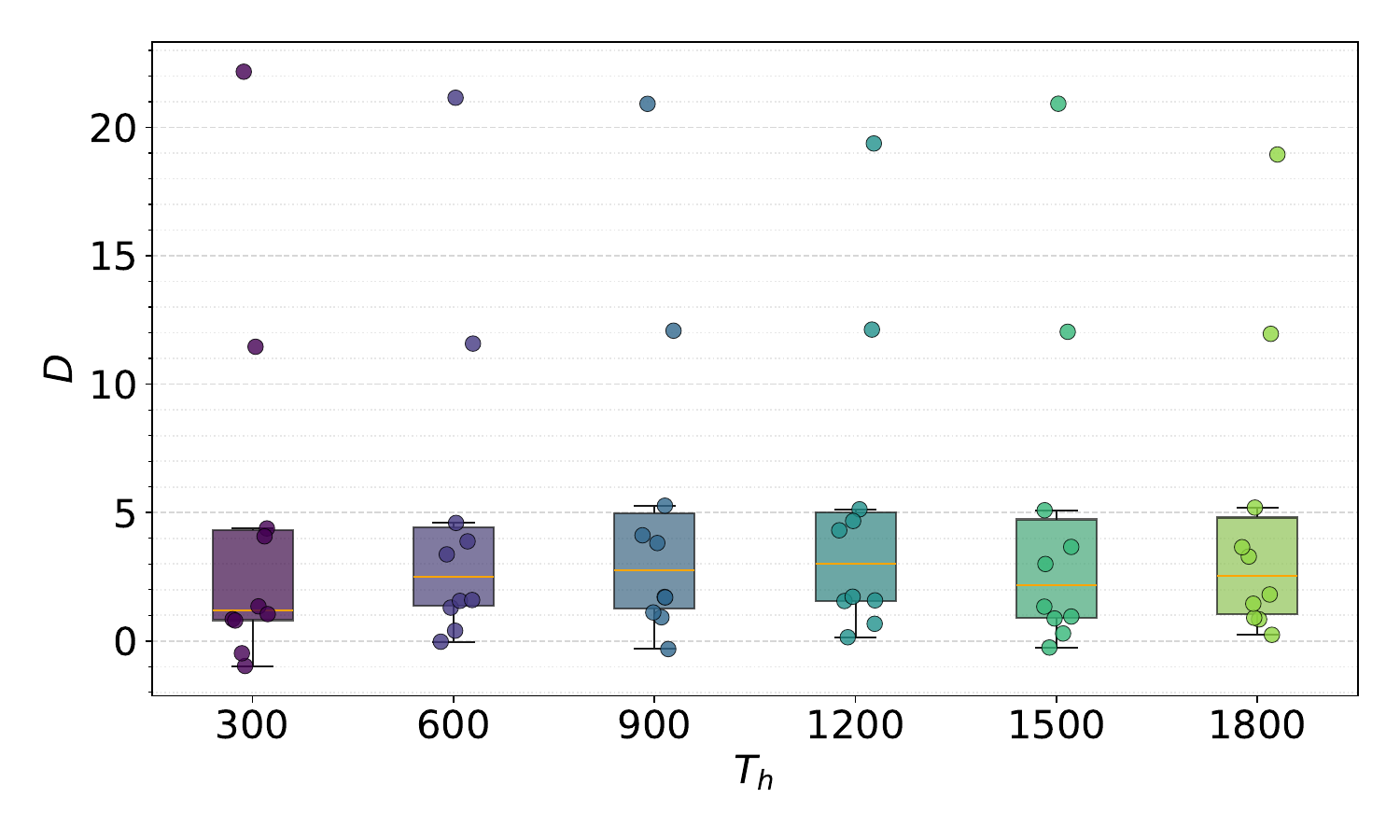}
    \caption{Distribution of percentage improvement over the centralised control system for different time horizons ($T_h$). Each box contains 10 points (one per perturbation) computed as the total improvement over the centralised control at the end of the operational day.}
    \label{fig:improve}
\end{figure}

Figure~\ref{fig:improve} illustrates the distribution of the percentage improvement ($D$) achieved by the SO-TMS with respect to the centralised control across the various time horizons considered. For each perturbed scenario, the total delay under a given control strategy is calculated as the sum of the delays of all trains upon exiting the control area. Consequently, the percentage improvement $D$ is defined as the relative variation in total delay provided by the SO-TMS for each scenario, formulated as:\begin{equation*}
    D = \frac{d_c - d_{so}}{d_c} \times 100
\end{equation*}where $d_c$ and $d_{so}$ represent the total accumulated delay at the exit of the control area for the centralised and SO-TMS approaches, respectively. Each distribution presented in the figure encompasses the $D$ values resulting from the ten perturbed scenarios, evaluated at the end of the operational day.

The results indicate comparable average performance across the different configurations, with an average improvement of approximately 5\% over the centralised approach. Shorter horizons exhibit slightly greater variability, with peak improvements exceeding 22\% alongside minimum results approaching -1\%. Overall, these results suggest that the time horizon does not significantly affect performance. Because the impact on overall delay does not uniquely delineate a preferred value for $T_h$, the subsequent analysis examines the internal metrics of the individual modules, beginning with the modifications induced on the social interaction graph topology.

\begin{figure}[ht]
    \centering
    \includegraphics[width=1\linewidth]{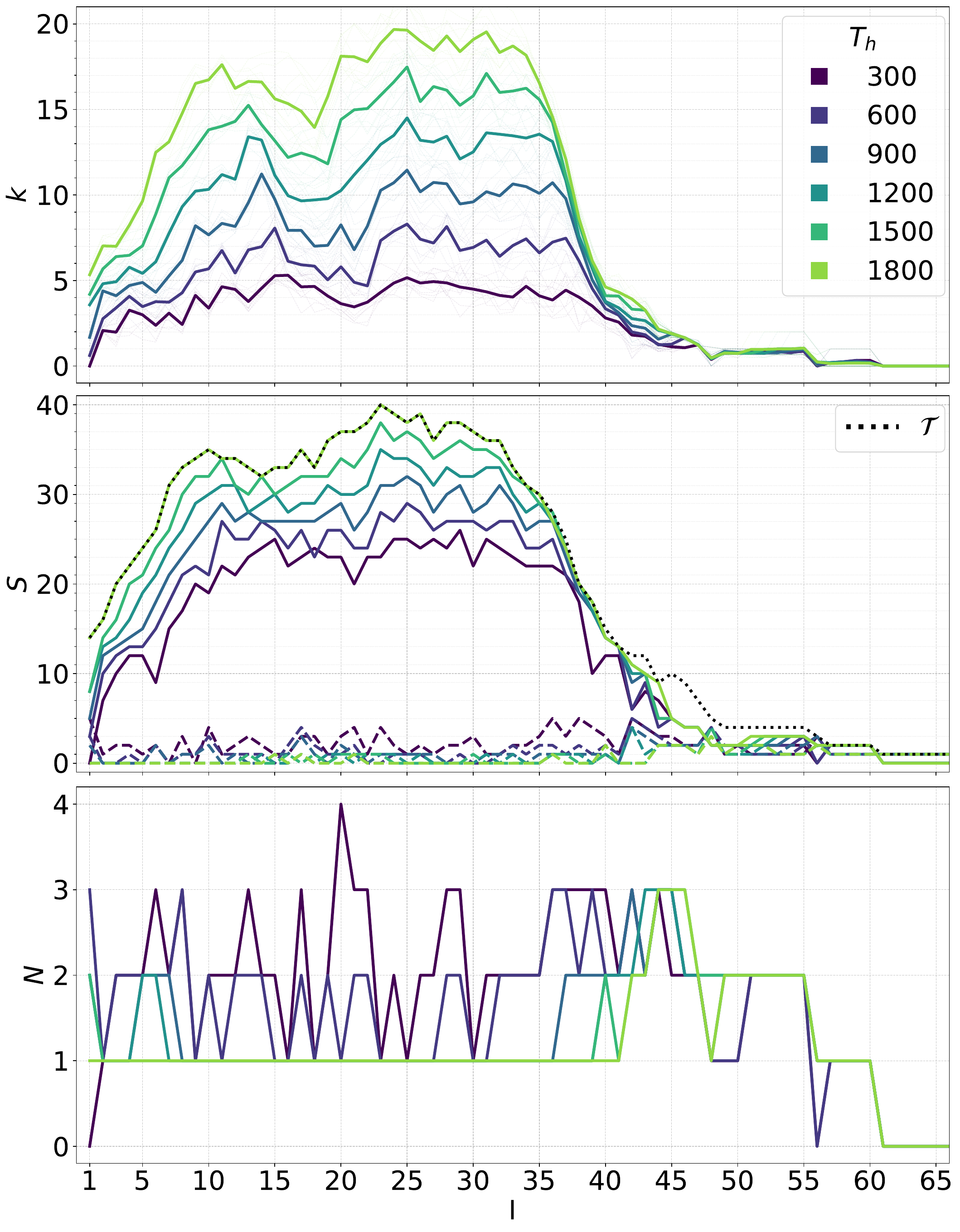}
    \caption{Topological data of the social interaction graph. \textbf{Top:} average degree of the graph across iterations, corresponding to the average neighbourhood size. \textbf{Middle:} maximum size of the connected components across perturbations excluding isolated trains (solid coloured lines) and the maximum number of isolated trains across perturbations (dashed lines). For reference, the total number of trains in the railway network is shown as a dotted black line. \textbf{Bottom:} maximum number of connected components across perturbations at each iteration, excluding isolated trains.}
    \label{fig:cc_plot}
\end{figure}

Figure~\ref{fig:cc_plot} illustrates the topological features of the social interaction graph $G_{I}$. As anticipated by the inclusion criterion, extending the horizon leads to an increase in neighbourhood size, as measured by the average node degree $k$ (see top panel in Figure~\ref{fig:cc_plot}). Consequently, the social interaction graph becomes more connected, resulting in a fewer number of larger connected components (see middle and bottom panels in Figure~\ref{fig:cc_plot}). The middle panel reports the size $S$ of the largest connected component across perturbations (solid lines) as well as the maximum number of isolated trains (dashed lines). The total number of active trains in the network $\mathcal{T}$ is also shown for reference (dotted black line). The bottom panel shows the maximum number of connected components $N$ across perturbations as $T_h$ varies, excluding isolated trains from the calculations.

Data indicates that adopting larger time horizons leads to the formation of a highly connected graph. Although each neighbourhood only includes directly conflicting nodes, high values of $T_h$ increase the average degree of the graph.
Consequently, the size of the main connected component grows, and it includes the entirety of active trains when $T_h=\unit[1800]{s}$. Conversely, larger time horizons reduce the presence of isolated trains. Also, the number of connected components decreases to a single giant component when $T_h$ is high, while the social interaction graph is often disconnected when $T_h\leq\unit[600]{s}$. These measures also change during the course of the simulated day, owing to the evolution of the traffic. During the first iterations, trains enter the network and the social interaction graph is not connected. Neighbourhoods increase in size, and so does the connectivity within the graph (lower number of larger connected components). The last iterations are usually characterised by low traffic, with trains leaving the infrastructure, and a larger number of connected components found at its exits. The formation of a single, large connected component can influence the dynamics of the consensus phase, owing to the variety and compatibility of the generated local solutions.

\begin{figure}[ht]
    \centering
    \includegraphics[width=1\linewidth]{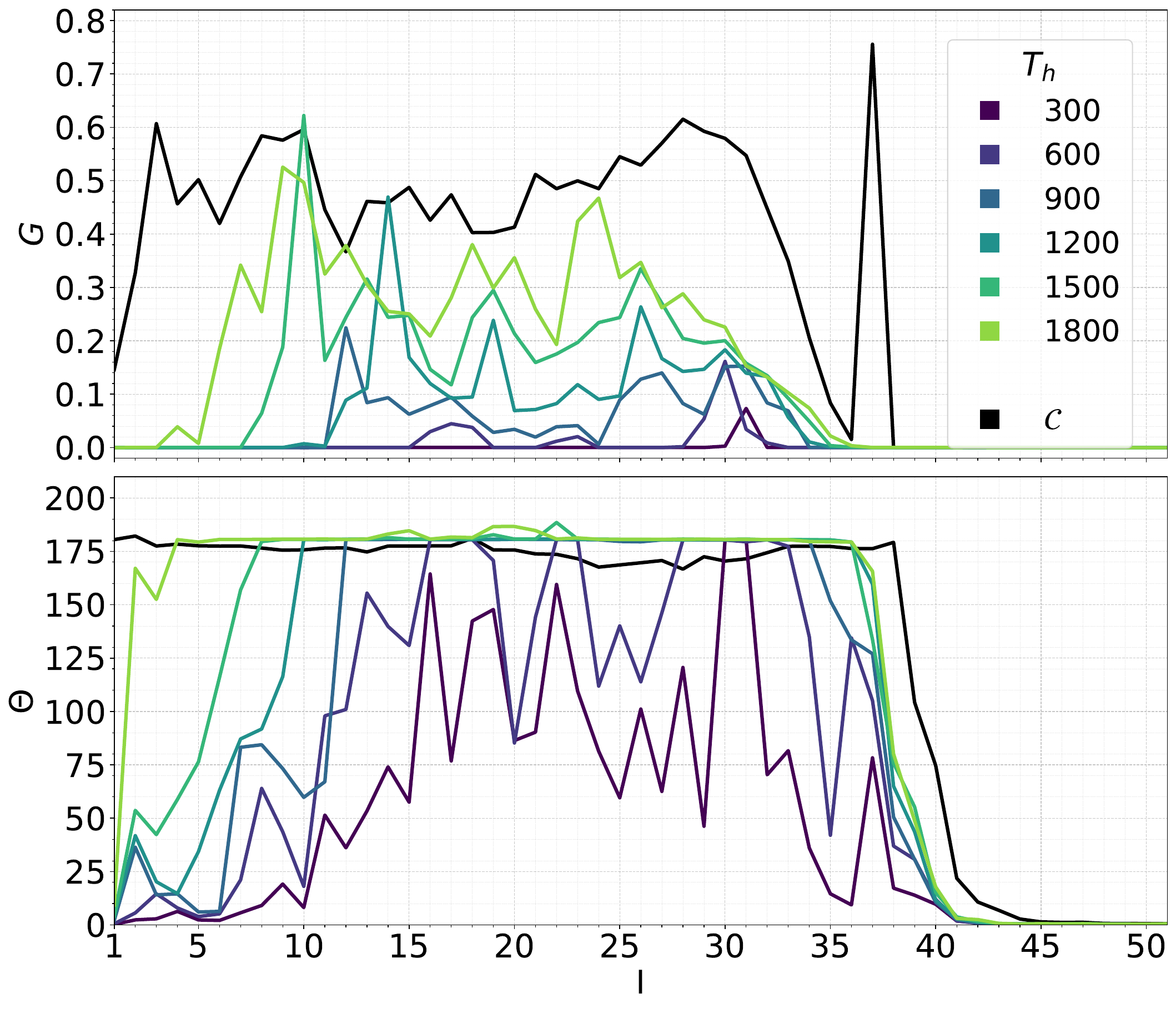}
    \caption{Performance of the hypothesis generation module. \textbf{Top:} maximum optimality gap $G$ across all perturbation scenarios for both the decentralised control (maximum gap among trains) and the centralised reference ($\mathcal{C}$). \textbf{Bottom:} maximum amount of computation time ($\Theta$, in seconds) necessary to conclude the hypothesis generation across all perturbations, alongside the maximum time for the centralised reference.}
    \label{fig:gen_gap_time}
\end{figure}

Large neighbourhoods result in a higher computational complexity for the local MILP problems. Figure~\ref{fig:gen_gap_time} displays, in the upper panel, the maximum optimality gap $G$ of local solutions across iterations, compared to the gap of the centralised solver $\mathcal{C}$. The gap, expressed as a fraction, indicates the distance from the optimal solution. To highlight the worst-case performance, the value for the centralised solver represents the maximum gap found at each iteration across the ten perturbed scenarios. Similarly, for the decentralised control, it reports the absolute maximum, across all scenarios, of the gap recorded among all trains at each iteration. The bottom panel shows the maximum time ($\Theta$, in seconds) required by the trains to complete the generation of hypotheses, compared to the peak time taken by the centralised approach to compute the new schedule. The data for both the centralised and decentralised systems represents the maximum values recorded across all perturbed scenarios at each iteration, with the decentralised data specifically reflecting the longest computation time among all active trains.

The results highlight that an increase in the time horizon is associated with an increased distance from the optimal solution in these worst-case scenarios. The adoption of short horizons allows for the selection of near-optimal solutions at each stage---with $T_h = 300$ maintaining maximum gaps of just $0.073$---while the highest values of $T_h$ demonstrate larger gaps. Specifically, the decentralised control reaches an absolute maximum gap of $0.622$ (at $T_h = 1500$), approaching the worst-case performance of the centralised control, which reaches a peak overall gap of $0.756$. Computation times confirm that introducing a larger number of variables increases the computational burden. During solution generation, trains inside and outside the neighbourhood are treated differently: the routes of the former can be substantially modified (re-routing), while those of the latter can only be adjusted by shifting the infrastructure utilisation (re-timing). Consequently, broader neighbourhoods entail longer computation times. For the decentralised case, the absolute maximum computation time peaks at $188.52$ seconds (at $T_h = 1500$), thereby surpassing the maximum peak time of $182.18$ seconds required by the centralised solver.

\begin{figure}[ht]
    \centering
    \includegraphics[width=1\linewidth]{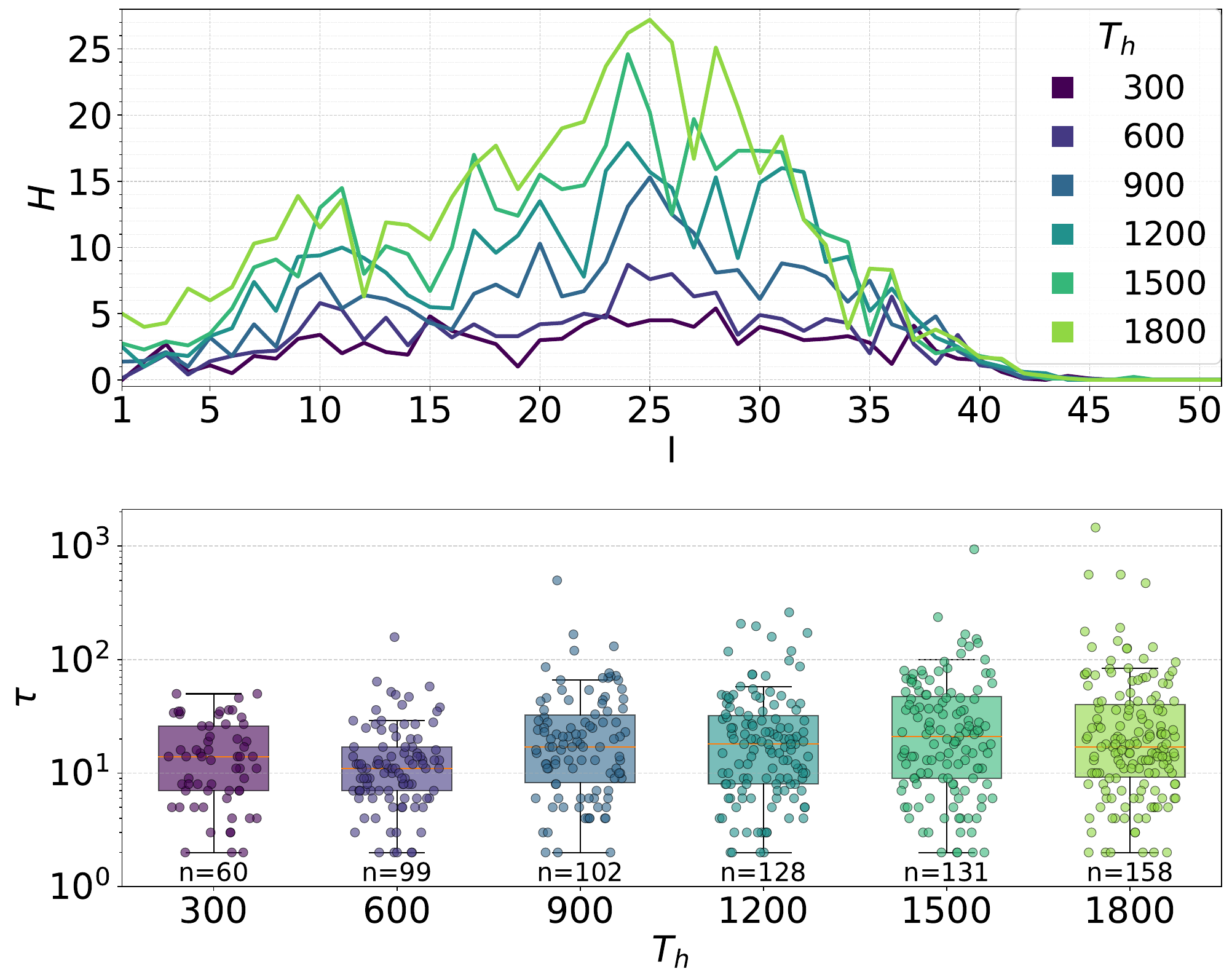}
    \caption{Performance of the sharing and consensus phases. \textbf{Top:} average increment of unique hypotheses ($H$) retained by trains after sharing. \textbf{Bottom:} distribution of the iterations ($\tau$) required to reach consensus, plotted on a logarithmic scale. The values $n$ explicitly indicate the count of processes that required active negotiation, filtering out isolated trains or trivially compatible initial states ($\tau=1$).}
    \label{fig:hp_share_increment}
\end{figure}

The hypothesis sharing phase is intended to increase the diversity and number of compatible solutions within the neighbourhoods. In Figure~\ref{fig:hp_share_increment}, the top panel displays the average increment $H$ in the number of unique solutions retained by trains after sharing. In the tests conducted, each train generates a maximum of two hypotheses: a newly computed schedule and a fail-safe option corresponding to the previous RTTP. Because sharing is restricted to direct neighbours and the generation itself considers only local interactions, the propagation of information within the network remains circumscribed. However, the graph clearly illustrates a stratification based on the time horizon: larger values of $T_h$ lead to a considerable increase in $H$. This occurs because a broader neighbourhood allows trains to intercept and accumulate a significantly larger pool of alternatives, which must then be evaluated during the hypothesis selection phase.

While local agreement is ultimately guaranteed in all connected components at every iteration, the complexity of the consensus process displays profound variations as the time horizon changes. The bottom panel of Figure~\ref{fig:hp_share_increment} isolates only the processes that required active negotiation (i.e., instances where initial hypotheses were not immediately compatible). The distribution, plotted on a logarithmic scale, highlights two major consequences of expanding $T_h$. First, the frequency of actual conflicts increases, with the number of necessary negotiation processes ($n$) rising steadily from $60$ at $T_h = 300$ to $158$ at $T_h = 1800$. Second, although the vast majority of conflicts are resolved swiftly, broader horizons introduce severe worst-case scenarios. The increased number of interdependent agent pairs induces cascading effects during negotiation, extending the maximum number of iterations ($\tau$) required to converge. For the highest values of $T_h$, this effect pushes the convergence time up to $10^3$ steps, aligning with the theoretical complexities of consensus processes predicted in~\cite{damato2020towards}.

\begin{figure}[ht]
    \centering
    \includegraphics[width=1\linewidth]{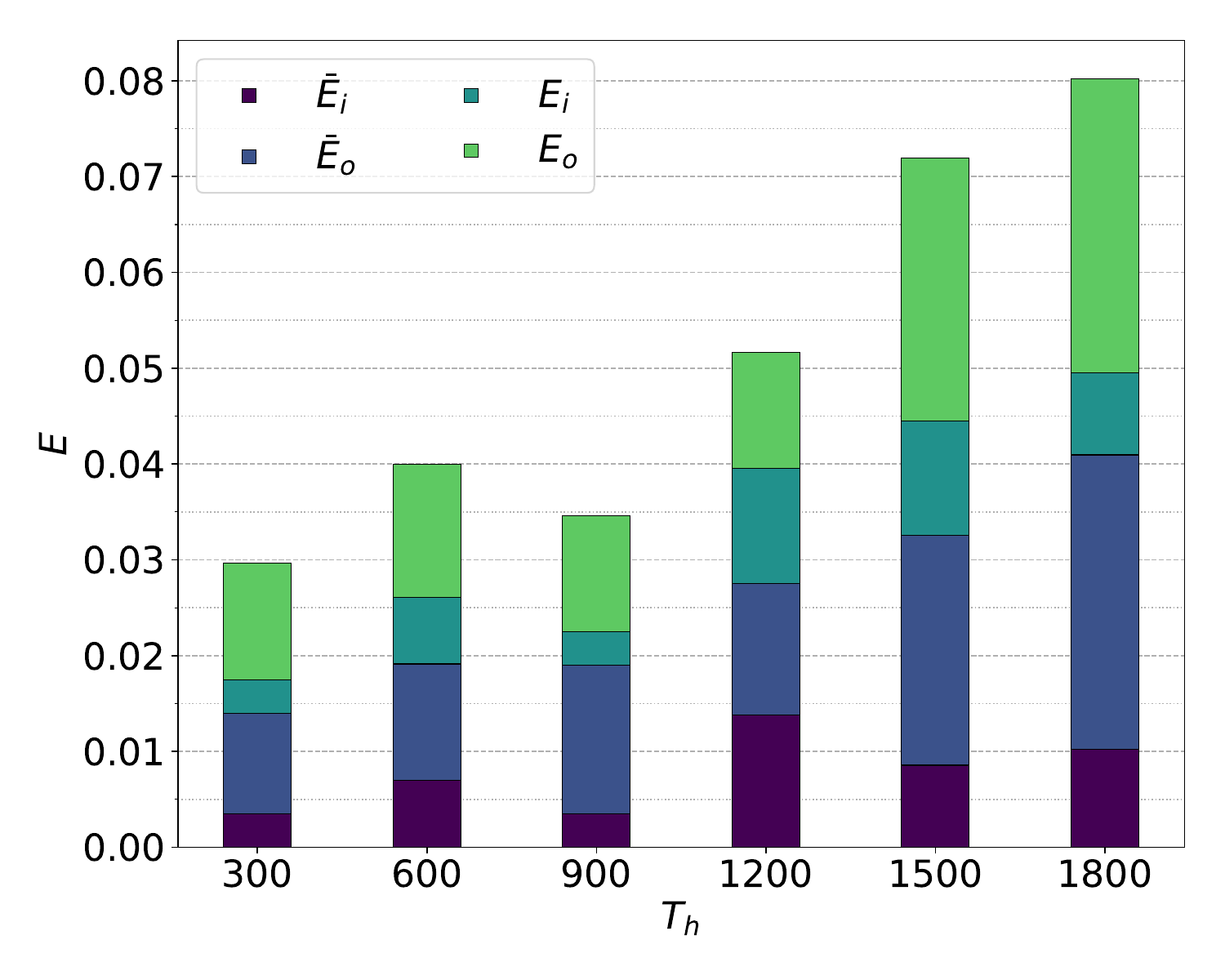}
    \caption{Total count $E$ of incompatible train pairs detected at the end of the merge phase for different values of the time horizon $T_h$. Incompatibilities resolved via simple shifts ($E_i$ and $E_o$) are distinguished from those requiring solver intervention ($\bar{E}_i$ and $\bar{E}_o$), further classified into \emph{in-area} and \emph{out-of-area} categories based on the operational state of the involved trains.}
    \label{fig:conflicts}
\end{figure}

In the final stage, the merge of new schedules produces a global RTTP that is subsequently validated by a central unit to guarantee feasibility. Residual incompatibilities do not necessarily require the entire plan to be rewritten, as they can often be resolved through simple re-timing of the routes; however, in certain cases, corrective intervention via the MILP solver is necessary. The Figure~\ref{fig:conflicts} illustrates the frequency $E$ of incompatible train pairs detected at the end of the merge phase. The anlysis distinguishes between conflicts resolved via simple shifts ($E_i$ and $E_o$) and those requiring solver intervention($\bar{E}_i$ and $\bar{E}_o$). Incompatibilities are divided into two categories: \emph{in-area} ($E_i$ and $\bar{E}_i$), which includes pairs of trains currently running on the infrastructure (including isolated trains), and \emph{out-of-area} ($E_o$ and $\bar{E}_o$), relating to pairs where one train is active and the other is waiting to access the network. The data indicates that repair interventions remain infrequent, occurring at most 4\% of all tests. Considering also incompatibilities solved without central repair, the total frequency of detected anomalies reaches the maximum value of 8\%. Nevertheless, a positive correlation is visible between the time horizon $T_h$ and the occurrence of repairs. This trend mirrors the increase in the optimality gap observed before, suggesting that expanding the local decision scope does not necessarely improve the global plan coherence or reduce the demand for central coordination.

The total number of incompatible pairs is small compared to the overall volume of tests, though Figure~\ref{fig:conflicts} demonstrates that the frequency of these conflicts increases as the time horizon $T_h$ expands. The majority of these conflicts involve trains currently out-of-area (waiting to access the network), while a smaller proportion affects trains simultaneously active on the infrastructure. This trend is primarily attributable to the exclusion from the decision-making process of trains without neighbours---concerning out-of-area and isolated in-area trains---whose routes remain unmodified by the negotiating trains. Incompatibilities are also recorded between active trains belonging to the same connected component, often as a consequence of the hypothesis sharing phase or because local hypothesis generation only partially accounts for long-range constraints. Crucially, conflicts also frequently emerge between trains in distinct connected components or involving isolated trains. In these cases, conflicts arise because trains in disconnected components optimise their schedules independently, without any mutual awareness or coordination. Consequently, while these plans are locally feasible, they may be globally inconsistent, leading to conflicts that only emerge during the final merge at the traffic control centre.

\section{Discussion and Conclusions}\label{sec:concl}
In this study, we analysed the impact of varying the predictive neighbourhood horizon $T_h$ within a decentralised railway traffic management architecture. The experimental results identify a structural trade-off associated with the choice of this parameter. While extending the horizon enlarges neighbourhoods and increases the set of shared hypotheses, it concurrently raises the computational burden of local MILP problems, reduces local solution quality under fixed computational budgets, and increases the incidence of incompatibilities during the central merge phase. This last effect stems primarily from a mismatch between the scope of local decisions and the unchanged schedule entries of trains that do not participate in the negotiation process, such as isolated convoys or those yet to enter the network. Consequently, broader horizons facilitate more extensive local planning but simultaneously increase the probability that these local modifications conflict in establishing a global plan. Notably, the total network delay did not show systematic improvements with larger horizons, indicating that an extended horizon does not strictly correlate with better system performance in decentralised settings. This is surprising because small horizons should correspond to myopic, greedy decisions. However, the closed-loop simulation demonstrates that possible myopic decisions can be easily corrected by downstream replanning. Overall, the analysis suggests that a short time horizon is preferable.

\printbibliography

@article{corman2015review,
  title={A review of online dynamic models and algorithms for railway traffic management},
  author={Corman, Francesco and Meng, Ling},
  journal={IEEE Transactions on Intelligent Transportation Systems},
  volume={16},
  number={3},
  pages={1274--1284},
  year={2015},
  publisher={IEEE}
}

@article{cacchiani2014overview,
  title={An overview of recovery models and algorithms for real-time railway rescheduling},
  author={Cacchiani, Valentina and Huisman, Dennis and Kidd, Mark and Kroon, Leo and Toth, Paolo and Veelenturf, Lucas and Wagenaar, Jaap},
  journal={Transportation Research Part B: Methodological},
  volume={63},
  pages={15--37},
  year={2014},
  publisher={Elsevier}
}

@article{fang2015survey,
  title={A survey on problem models and solution approaches to rescheduling in railway networks},
  author={Fang, Wei and Yang, Shengxiang and Yao, Xin},
  journal={IEEE Transactions on Intelligent Transportation Systems},
  volume={16},
  number={6},
  pages={2997--3016},
  year={2015},
  publisher={IEEE}
}

@article{zhang2020fast,
  title={A fast approach for reoptimization of railway train platforming in case of train delays},
  author={Zhang, Yongxiang and Zhong, Qingwei and Yin, Yong and Yan, Xu and Peng, Qiyuan},
  journal={Journal of advanced transportation},
  volume={2020},
  number={1},
  pages={5609524},
  year={2020},
  publisher={Wiley Online Library}
}

@article{ghasempour2020adaptive,
  title={Adaptive railway traffic control using approximate dynamic programming},
  author={Ghasempour, Taha and Heydecker, Benjamin},
  journal={Transportation Research Part C: Emerging Technologies},
  volume={113},
  pages={91--107},
  year={2020},
  publisher={Elsevier}
}

@inproceedings{shang2018distributed,
  title={Distributed model predictive control for train regulation in urban metro transportation},
  author={Shang, Fei and Zhan, Jingyuan and Chen, Yangzhou},
  booktitle={2018 21st International Conference on Intelligent Transportation Systems (ITSC)},
  pages={1592--1597},
  year={2018},
  organization={IEEE}
}

@inproceedings{yong2017swarm,
  title={Decentralized, autonomous train dispatching using swarm intelligence in railway operations and control},
  author={Yong, Chen and Ullrich, Michael and Jiajian, Li},
  booktitle={7th International Conference on Railway Operations Modelling and Analysis (RailLille2017)},
  pages={521--540},
  year={2017},
  organization={IAROR}
}

@article{vanthielen2019conflict,
  title={Towards a conflict prevention strategy applicable for real-time railway traffic management},
  author={Van Thielen, Stijn and Corman, Francesco and Vansteenwegen, Pieter},
  journal={Journal of Rail Transport Planning \& Management},
  volume={11},
  pages={100139},
  year={2019},
  publisher={Elsevier}
}

@article{damato2020towards,
  title={Towards self-organizing railway traffic management: concept and framework},
  author={D'Amato, Leo and Naldini, Federico and Tibaldo, Valentino and Trianni, Vito and Pellegrini, Paola},
  journal={Journal of Rail Transport Planning \& Management},
  volume={29},
  pages={100427},
  year={2020},
  publisher={Elsevier}
}

@article{naldini2026self,
  title={Self-Organizing Railway Traffic Management},
  author={Naldini, Federico and Oddi, Fabio and d'Amato, Leo and Marli{\`e}re, Gr{\'e}gory and Trianni, Vito and Pellegrini, Paola},
  journal={arXiv preprint arXiv:2601.17017},
  year={2026}
}

@article{pellegrini2015recife,
  title={RECIFE-MILP: an effective MILP-based heuristic for the real-time railway traffic management problem},
  author={Pellegrini, Paola and Marli{\`e}re, Gr{\'e}gory and Pesenti, Raffaele and Rodriguez, Joaquim},
  journal={IEEE Transactions on Intelligent Transportation Systems},
  volume={16},
  number={5},
  pages={2609--2619},
  year={2015},
  publisher={IEEE}
}

@article{detrain2006self,
  title={Self-organized structures in a superorganism: do ants “behave” like molecules?},
  author={Detrain, Claire and Deneubourg, Jean-Louis},
  journal={Physics of Life Reviews},
  volume={3},
  number={3},
  pages={162--187},
  year={2006},
  publisher={Elsevier}
}

@article{seeley2012stop,
  title={Stop signals provide cross inhibition in collective decision-making by honeybee swarms},
  author={Seeley, Thomas D and Visscher, P Kirk and Schlegel, Thomas and Hogan, Patrick M and Franks, Nigel R and Marshall, James AR},
  journal={Science},
  volume={335},
  number={6064},
  pages={108--111},
  year={2012},
  publisher={American Association for the Advancement of Science}
}

@article{dorigo2021swarm,
  title={Swarm robotics: past, present, and future},
  author={Dorigo, Marco and Theraulaz, Guy and Trianni, Vito},
  journal={Proceedings of the IEEE},
  volume={109},
  number={7},
  pages={1152--1165},
  year={2021},
  publisher={IEEE}
}

@article{liu2023cognitive,
  title={Cognitive network architecture systems to provide intelligent services: An intelligent self-organization approach with a game-based incentive mechanism},
  author={Liu, Yang and Gui, Jing and Xiong, Naixue},
  journal={IEEE Systems, Man, and Cybernetics Magazine},
  volume={9},
  number={1},
  pages={25--36},
  year={2023},
  publisher={IEEE}
}

@article{arellanes2021self,
  title={Self-organizing software models for the internet of things: complex software structures that emerge without a central controller},
  author={Arellanes, Damian},
  journal={IEEE Systems, Man, and Cybernetics Magazine},
  volume={7},
  number={3},
  pages={4--9},
  year={2021},
  publisher={IEEE}
}

@article{strombom2018self,
  title={Self-organized traffic via priority rules in leaf-cutting ants},
  author={Str{\"o}mbom, Daniel and Dussutour, Audrey},
  journal={PLOS Computational Biology},
  volume={14},
  number={10},
  pages={1--13},
  year={2018},
  publisher={Public Library of Science}
}

@article{moussaid2012traffic,
  title={Traffic instabilities in self-organized pedestrian crowds},
  author={Moussa{\"i}d, Mehdi and Guillot, Elisa G and Moreau, Mathieu and Fehrenbach, Jonas and Chabiron, Olivier and Lemercier, Samuel and Pettr{\'e}, Julien and Appert-Rolland, C{\'e}cile and Degond, Pierre and Theraulaz, Guy},
  journal={PLOS Computational Biology},
  volume={8},
  number={3},
  pages={1--10},
  year={2012},
  publisher={Public Library of Science}
}

@article{murakami2021mutual,
  title={Mutual anticipation can contribute to self-organization in human crowds},
  author={Murakami, Hisashi and Feliciani, Claudio and Nishiyama, Yuta and Nishinari, Katsuhiro},
  journal={Science Advances},
  volume={7},
  number={12},
  pages={eabe7758},
  year={2021},
  publisher={American Association for the Advancement of Science}
}

@incollection{helbing2005self,
  title={Self-organized control of irregular or perturbed network traffic},
  author={Helbing, Dirk and L{\"a}mmer, Stefan and Lebacque, Jean-Patrick},
  booktitle={Managing Complexity},
  pages={239--274},
  year={2005},
  publisher={Springer}
}

@incollection{gershenson2012self,
  title={Self-organizing urban transportation systems},
  author={Gershenson, Carlos},
  booktitle={Complexity in Engineering},
  pages={269--279},
  year={2012},
  publisher={Springer}
}

@article{bucchiarone2020agent,
  title={Agent-based framework for self-organization of collective and autonomous shuttle fleets},
  author={Bucchiarone, Antonio and De Sanctis, Martina and Bencomo, Nelly},
  journal={IEEE Transactions on Intelligent Transportation Systems},
  volume={22},
  number={6},
  pages={3631--3643},
  year={2020},
  publisher={IEEE}
}

@article{gerrits2024self,
  title={Towards self-organizing logistics in transportation: a literature review and typology},
  author={Gerrits, Berry and van Heeswijk, Wouter and Mes, Martijn},
  journal={International Transactions in Operational Research},
  volume={31},
  number={3},
  pages={1309--1374},
  year={2024},
  publisher={Wiley Online Library}
}

@article{mittal2024efficient,
  title={Efficient self-organization of informal public transport networks},
  author={Mittal, Kanupriya and Timme, Marc and Schr{\"o}der, Malte},
  journal={Nature Communications},
  volume={15},
  number={1},
  pages={4910},
  year={2024},
  publisher={Nature Publishing Group}
}

@article{khadilkar2018scalable,
  title={A scalable reinforcement learning algorithm for scheduling railway lines},
  author={Khadilkar, Harshad},
  journal={IEEE Transactions on Intelligent Transportation Systems},
  volume={20},
  number={2},
  pages={727--736},
  year={2018},
  publisher={IEEE}
}

@article{nash2004railroad,
  title={Railroad simulation using OpenTrack},
  author={Nash, Andrew and Huerlimann, Daniel},
  journal={WIT Transactions on The Built Environment},
  volume={74},
  year={2004},
  publisher={WIT Press}
}
\end{document}